\documentstyle[12pt,aps,epsf,preprint,tighten]{revtex} 
\begin{document} \draft 
\title{ \begin{flushright}
{\small Preprint HKBU-CNS-9813\\ April 1998}
\end{flushright} Analytical Results For The Steady State Of Traffic Flow
 Models With Stochastic Delay } 

\author{  Bing-Hong Wang$^{[1, 2, 3]}$, Lei Wang$^{[2]}$,
P. M. Hui$^{[4]}$ and Bambi Hu$^{[1, 5]}$ }  

\address{ $^{[1]}$ Department of Physics and Center for Nonlinear Studies,\\
Hong Kong Baptist University, Hong Kong, China \\ 
$^{[2]}$ Department of Modern Physics and Nonlinear Science Center,\\
 University of Science and Technology of China, Hefei 230026, China\\ 
$^{[3]}$ CCAST (World Laboratory), P.O.Box 8730, Beijing 100080, China\\
$^{[4]}$ Department of Physics, The Chinese University of Hong Kong,\\
 Shatin, New Territories, Hong Kong, China\\
$^{[5]}$ Department of Physics, University of Houston, Houston TX 77204, USA} 


\date{\today}  
\maketitle 
 
\begin{abstract} 
Exact mean field equations are derived analytically to
give the fundamental diagrams, i.e., the average speed - car 
density relations, for the Fukui-Ishibashi one-dimensional
traffic flow cellular automaton model of high speed vehicles
$\left(v_{max}=M>1\right) $ with stochastic delay. Starting with the basic 
equation describing the time evolution of the number of empty sites in front 
of each car, the concepts of inter-car spacings longer and shorter than $M$ are
introduced.  The probabilities of having long and short spacings on the road 
are calculated.  For high car densities $(\rho \geq 1/M)$, it is shown that 
inter-car spacings longer than $M$ will be shortened as the traffic flow 
evolves in time, and any initial configurations approach a steady state in 
which all the inter-car spacings are of the short type. Similarly for 
low car densities $(\rho \leq 1/M)$, it can be shown that traffic flow 
approaches an asymptotic steady state in which all the inter-car spacings are 
longer than $M-2$.  The average traffic speed is then obtained analytically 
as a function of car density in the asymptotic steady state. The fundamental
diagram so obtained is in excellent agreement with simulation data.
\end{abstract}  

\vspace*{0.15 true in}
\pacs{PACS numbers:  64.60.Ak, 05.40.+j, 05.70.Jk, 89.40.+k } 

\begin{center}
{\bf I. INTRODUCTION}
\end{center}

Recently, there has been much interest in studying traffic flow problems 
within the context of cellular automaton (CA) models [1-4].  Compared with
the fluid dynamical approaches to traffic flow problems, the CA models are 
conceptually simpler, and can be readily implemented on computers.  
These models capture the complexity of the nonlinear character of the problem 
and provide clear physical pictures [2,5-7].  For example, CA models show 
the existence of a transition between a moving phase and a jamming phase 
in the traffic of a city as the car density is varied [4].  These models also 
have the advantages that they can be easily modified to deal with the effects
of different kinds of realistic conditions, such as road blocks and
hindrances, traffic accidents [8], highway junctions [9],  
overpasses [10], vehicle acceleration [11],
quenched disorderness [12], 
stochastic delay due to driver's reactions [13], 
anisotropy of car distributions in different driving directions [14],
faulty traffic lights [15], etc.  Recently, CA models 
have been successfully applied to study traffic flow in a city by 
performing high speed simulations on the actual road map of the city of 
Dallas [16].
In view of the increasing importance of CA models in studying traffic 
flow problems, it is thus important to understand these models in more 
detail, especially from the point of view of statistical mechanics and 
nonlinear dynamics. 

The basic one dimensional (1D) CA model for highway traffic flow [3,13]  
is the CA rule
184 [1].  This model describes single-lane traffic on a road of
length $L$ with periodic boundary condition. Each of the $L$ sites can either be
empty or occupied by one vehicle. Let $N$ be total number of cars, then the
average vehicle density on the road is $\rho =N/L$.  The cars move
from the left to the right according to the following rules.  
All the cars attempt a move of one step to the right simultaneously
at each time step.  If the site in front of a car is not occupied at that  
time step, the car moves one site ahead.  
Otherwise, the car cannot move. 
This simple model predicts a transition
from laminar traffic flow to start-stop wave as the car density increases.

There are many variations on the basic model. 
Nagel and Schreckenberg (NS) considered the effects
of acceleration and stochastic delay of vehicles with high speed [3,13]. 
In the NS model, a car can move at most by $M$ sites in a time step.  The 
actual speed at a time step depends on the spacing in front.  If the speed 
in the present time step is less $M$ and the spacing ahead allows, then 
the speed increases by one unit in the next time step.  If the spacing 
ahead is less than the speed in the present time step, then the speed 
is reduced to the value allowed by the spacing, and thus leads 
to a deceleration.  In addition, there is a probability that the speed of a 
car is reduced by one unit in the next time step.  Thus the NS model 
captures the features of gradual acceleration, deceleration and 
randomization in realistic traffic flows. 

Fukui and Ishibashi (FI) introduced another variation on the basic 
model [17] in which the cars can move by at most $M$ sites in one time 
step if they are not blocked by cars in front.  More precisely, 
if the number of empty sites $C$ in front of a car is larger than 
$M$ at time $t$, then it can move forward $M$ ($M-1$) sites in the 
next time step with probability $1-f$ ($f$).  Here, the probability $f$
represents the degree of stochastic delay.  The $f=0$ model is referred 
to as the deterministic FI model, while the $f=1$ case is the deterministic
FI model with the maximum speed $v_{max} = M-1$.  If $C < M$ at time $t$, 
then the car can only move by $C$ sites in the next time step.  The FI 
model differs from the NS model in that the increase in speed may not 
be gradual and stochastic delay only applies to the high speed cars.  
Obviously, the two models are identical for $M=1$. 

Fukui and Ishibashi have performed numerical simulations on the 
model [17].  The focus in traffic flow problems is the so called 
fundamental diagram, which is the relationship between the average 
speed in the steady state and the car density.  While numerical 
simulations provide us with accurate fundamenetal diagrams, it is 
desirable and useful to have a better qualitative understanding of
the numerical results within some analytic approaches such as 
mean field theories.  Numerical studies and mean field theories have been 
extremely useful in providing detailed understanding in phase transitions
and critical phenomena in equilibrium statistical mechanics, and we 
foresee that they will be equally useful in the study of dynamical systems
such as the CA traffic flow models. 

Various mean field theories have been proposed for traffic flow 
models in 2D [10,14,19-21] and 1D [13,17,22-30].  In 1D, mean 
field approaches giving results in exact agreement with simulations 
have been given for the NS model [13,22] with $M=1$ and for 
the deterministic FI model [17].  In 2D, mean field theories have 
been proposed [10,14,20,21] for the model introduced 
by Biham {\em et al} [4] 
(BML model).  Most of the mean field approahces are macroscopic 
theories in that the consideration is based on the idea that the average 
duration 
that a car stays on a site, 
while depending on the speed of the cars, 
determines the blockage on the car behind it and thus, in turn, determines
the average speed.  

Recently, the steady state of CA traffic flow models have been studied
within statistical mechanical approaches [23-27].  While these 
studies are also mean field in nature, the approaches are based 
on {\em microscopic} consideration focusing on the time evolution of the 
occupancy on each site of the road.  A nonlinear mapping between 
the macroscopic average speeds at two consecutive time steps can 
then be derived by carrying out suitable statistical averages on the 
microscopic relations.  The stable fixed point of the mapping 
gives the steady state average speed as a function of car denisty.  For 
the deterministic FI model, results in exact agreement with numerical 
data have been obtained [26,27]; while for the FI model with delay, the 
microscopic approach gives results in good agreement with simulations.  
Such microscopic theory has the advantage that it provides a 
systematic approach for the derivation of mean field results for 
the steady state.  

An alternative microscopic approach based on the time evolution of 
inter-car spacings have also been proposed recently for the 
FI model [28-30].  The idea is similar to the 
car-oriented mean field theory (COMF) [22].  In Ref.[30], we 
studied the deterministic FI model for arbitrary $M$ and found that the 
inter-car spacings self-organize themselves into either long 
or short spacings in the steady state depending on the car density on 
the road.  The fundamental diagram so obtained is in exact agreement 
with simuations.  In this paper, we generalize this approach to study 
FI model with arbitrary $M$ and arbitrary degree of stochastic delay $f$.  
We are able to derive a general expression for the average car speed 
in the steady state as a function of $M$ and $f$ which is in excellent 
agreement with numerical data.  

The plan of the paper is as follows.  In Sec.II, we present the 
basic evolution equation for the inter-car spacings.  The concepts 
of long and short spacings are introduced.  The average speed is expressed 
in terms of the probabilities of finding long and short spacings.  
Section III gives a detailed discussion on how the inter-car spacings 
evolve in time according to the inter-car spacing of the car in front.  
Using the results in Sec.III, Sec. IV gives the steady state result in the 
high car density regime.  It is argued that all the spacings evolve into
short spacings in the steady state.  In Sec. V, we derive the average 
speed in the low density regime using the idea of detailed balance.  
Results are discussed in Sec. VI and compared with numerical data for 
$M=2$ and $M=3$ with arbitrary degree of stochastic delay.  The 
extension of the present approach to other traffic flow problems 
is also discussed. 

\vspace*{0.15 true in}
\begin{center}
{\bf II. THE DEPENDENCE OF AVERAGE TRAFFIC SPEED ON INTER-CAR SPACINGS}
\end{center}

Let $C_n(t)$ be the number of empty sites in front of the $n$-th car at
time $t$. It is also the distance between 
the $n$-th car and the ($n+1$)-th
car. The average distance between neighboring cars 
can be represented by 
$\bar C\equiv \frac{L-N}N=\frac 1\rho -1$. 
Let $v_n(t)$ be the number of sites 
that the $n$-th car moves during an update
at the time $t$, i.e. the update between time $t$ and $t+1$.  
The number of empty sites in
front of the $n$-th car at time $t+1$ is 
\begin{equation}
C_n(t+1)=C_n(t)+v_{n+1}(t)-v_n(t)
\end{equation}

Within the generalized FI traffic flow model with the 
maximum car velocity 
$v_{max}=M$ and a stochastic delay probability $f$,  the relationship  
between
the velocity of the $n$-th car and the inter-car spacing ahead 
at time $t$ is 
\begin{equation} 
v_n\left( t\right) =F_M\left( f,C_n\left( t\right) \right),
\end{equation}  
where
\begin{equation} 
    F_M(f,C)=
	\left\{\begin{array} {ll}
	    C, & \mbox{if }C\leq M-1\\
	    \left.
		\begin{array}{ll}
		M-1,\quad & \mbox{with probability }f\\ 
		M,\quad & \mbox{with probobility }1-f
		\end{array}
	    \right\} & \mbox{if }C \geq M.
	 \end{array}\right.
\end{equation}
The average speed at time $t$ is
\begin{eqnarray}   
V\left( t\right)=\frac 1N\sum\limits_{n=1}^Nv_n\left( t\right) 
=\frac 1N\left\{ \sum\limits_{C_n\left( t\right) \leq M-1}C_n\left(
t\right) +\sum\limits_{C_n\left( t\right) \geq M}(M-f)\right\}.
\end{eqnarray}
The sums in Eq.(4) correspond to sums over two different 
types of car-spacings.  An inter-car 
spacing is labelled a long spacing if it consists of $M$ or more sites, 
i.e. if $C_{n}(t) \geq M$; while an inter-car spacing is 
labelled a short spacing if it consists of $M-1$ or less sites, i.e. 
$C_{n}(t) \leq M-1$.
Let $N_m(t)$ be the number of cars at time $t$ with $m$ empty
sites ahead.  The probability that 
a car is found to have a spacing of $m$ sites ahead is give by
$P_{m}(t) = N_{m}(t)/N$.  
The number of long spacings $N_{long}(t)$ at time $t$ is given by 
$N_{long}\left( t\right) = \sum_{m\geq M} N_{m}(t)$, and  the number 
of short spacings $N_{short}(t)$ at time $t$ is given by 
$N_{short}\left( t\right) = \sum_{m=0}^{M-1} N_{m}(t)$.
The probability $P_{long}(t)$ of an inter-car spacing belonging to a long
spacing is 
$P_{long}\left( t\right) =N_{long}\left( t\right) /N$; while 
and the probability $P_{short}(t)$ of an inter-car spacing belonging 
to a short
spacing is
$P_{short}\left( t\right) =N_{short}\left( t\right) /N$.  
Thus, the average car speed in Eq.(4) 
can be expressed in terms of $P_{m}(t)$ as 
\begin{equation}
V\left( t\right) =\sum_{m=1}^{M-1} m \cdot P_{m}(t) + 
\left( M-f\right) P_{long}\left( t\right).  
\end{equation}

\vspace*{0.15 true in}
\begin{center}
{\bf III. TIME EVOLUTION OF INTER-CAR SPACINGS}
\end{center}

For the FI traffic flow model with stochastic delay, the 
inter-car spacings evolve in time in the following ways depending 
on whether the spacing is short or long {\em and} on the nature 
of the inter-car spacing of the car in front.  Suppose the 
spacing of the $n$-th car is {\em short}, i.e. 
$C_n(t)\leq M-1.$
If $C_{n+1}(t)\leq M-1$, then 
\begin{equation}
C_n(t+1)=C_{n+1}(t)<M-1.
\end{equation}
If $C_{n+1}(t)\geq M$, then
\begin{equation}
C_n(t+1)=\left\{\begin{array}{ll} M-1, \qquad & \mbox{with probability }f\\
				  M, \qquad & \mbox{with probability }1-f.
		 \end{array}\right.
\end{equation}
Equation (6) follows from Eqs.(1)-(3) that for $C_{n}(t) \leq M-1$
and $C_{n+1}(t) \leq M-1$, 
$C_n(t+1)=C_n(t)+C_{n+1}(t)-C_n(t)=C_{n+1}(t) \leq M-1.$
For Eq.(7), 
we introduce a stochastic Boolean
variable $\theta _n(f)$ describing the stochastic delay 
of the n-th car with
probability $f$:
\begin{equation}
\theta _n(f)\equiv \left\{\begin{array}{ll}
			1, \qquad & \mbox{with probability }f\\
			0, \qquad & \mbox{with probability }1-f.  
			  \end{array}\right.
\end{equation}
It then follows from Eqs.(1)-(3) that for $C_{n}(t) \leq M-1$  
and $C_{n+1}(t) \geq M$, 
\begin{equation}
C_n(t+1)=(M-1)\theta _{n+1}(f)+M\theta
_{n+1}(1-f), 
\end{equation}
and hence Eq.(7). 

Suppose the spacing of the $n$-th car is {\em long}, i.e. $C_n(t)\geq M$.
If $C_{n+1}(t)\leq M-1$, then
\begin{equation}
C_{n+1}(t)\leq C_n(t+1)\leq\left\{\begin{array}{ll} 
			   C_n(t) \qquad & \mbox{with probability }f\\
			   C_n(t)-1 \qquad & \mbox{with probability }1-f. 
			\end{array}\right.
\end{equation}
If $C_{n+1}(t)\geq M$, then
\begin{equation}
C_n(t+1)=C_n(t)+\left\{\begin{array} {ll}
		   1, \qquad & \mbox{with probability }f(1-f)\\
		   0, \qquad & \mbox{with pobobility }f^2+(1-f)^2\\
		  -1, \qquad & \mbox{with probability }f(1-f). 
		     \end{array}\right.
\end{equation}
The proof of Eq.(10) goes as follows.  It follows from Eqs.(1)-(3) 
that for $C_{n}(t) \geq M$ and $C_{n+1}(t) \leq M-1$, 
\begin{equation}
C_n(t+1)=C_n(t)+C_{n+1}(t)-[(M-1)\theta _n(f)+M\theta _n(1-f)].
\end{equation}
Since $C_n$(t)$\geq M$, $C_n(t+1)\geq C_{n+1}(t)$. For 
$C_{n+1}(t)\leq M-1$, 
$$C_{n+1}(t)-[(M-1)\theta _n(f )+M\theta _n(1-f )]\leq 
       \left\{\begin{array}{ll} 0, \qquad & \mbox{with probability }f\\
			       -1, \qquad & \mbox{with probability }1-f. 
	     \end{array}\right.$$
Hence Eq.(12) implies 
$$C_n(t+1)\leq\left\{\begin{array}{ll} 
	       C_n(t) \qquad & \mbox{with probability }f\\
	       C_n(t)-1 \qquad & \mbox{with probability }1-f, 
	       \end{array}\right.$$
and Eq.(10) is proven.

For Eq.(11), from Eqs.(1)-(3) and the conditions 
$C_n$(t)$\geq M$ and $C_{n+1}$(t)$\geq M$, we have:
\begin{eqnarray}
C_n(t+1)&=&C_n(t)+[(M-1)\theta _{n+1}(f)+M\theta _{n+1}(1-f)]-([(M-1)\theta
_n(f)+M\theta _n(1-f)])\nonumber\\
 &=&C_n(t)+[M\theta _{n+1}(1-f)-(M-1)\theta _n(f)]+[(M-1)\theta
   _{n+1}(f)-(M-1)\theta _n(f)]\nonumber\\
 &&+[M\theta _n(1-f)-M\theta _n(1-f)]+[(M-1)\theta
   _{n+1}(f)-M\theta _n(1-f)]\nonumber\\
 &=&C_n(t)+\left\{\begin{array}{ll} 
	1, \qquad &\vspace*{0.1 true in} \mbox{with probability }f(1-f) \\ 
	0, \qquad & \mbox{with probability }f^2+(1-f)^2 \\ 
       -1, \qquad & \mbox{with probability }f(1-f),\\
       \end{array}\right.
\end{eqnarray}
which is Eq.(11).  

Equations (6), (7), (10), and (11) give the time 
evolution of $C_{n}(t)$.  Coupled with Eq.(4) or Eq.(5) gives the 
time evolution of the average speed $V(t)$ and hence the asymptotic limit
can be studied. 

\vspace*{0.15 true in}
\begin{center}
{\bf IV. HIGH DENSITY CASE}
\end{center}

For high car densities ($\rho \geq 1/M$), the average 
inter-car spacings satisfies
$\bar C\equiv 1/\rho -1\leq M-1.$
It can be argued that in the asymptotic steady state of traffic flow, all
inter-car spacings become short spacings, i.e., 
\begin{equation}
C_n\left( t\right) \leq M-1,\qquad\forall n.
\end{equation}
>From Eq.(6), if every spacing is shorter than $M$, then 
\begin{equation}
C_n\left( t+1\right) =C_{n+1}\left( t \right), \qquad\forall n, 
\end{equation}
which implies that the spacing in front of the $n$-th car at $t+1$ is 
simply the spacing of the $n+1$-th car at time $t$.  Thus as 
time increases, the traffic evolves as a continuous shift 
in the numbering of cars.  Therefore, the situation corresponding to 
all the spacings are short is a steady state.  
Since
$$L-N=\sum\limits_nC_n\left( t\right) \leq \left( M-1\right) N,$$
the condition in Eq.(15) holds for $\rho \geq 1/M$.  Under this condition,
the average speed $V$ is simply the number of empty sites divided by the
total number of cars:
\begin{equation}
V=\frac{L-N}N=\frac 1\rho -1, \qquad\mbox{for $\rho \geq 1/M$}.
\end{equation}

The time evolution of $C_{n}(t)$ (Eqs.(6), (7), (10), (11)) ensures 
that for $\rho \geq 1/M$ the steady state in which all inter-car
spacings are short spacings is approached asymptotically {\em regardless}
of the initial state of the traffic flow.  The proof has been 
given [29,30]
for the deterministic FI model
with 
arbitrary $v_{max} = M$.  While a similar proof can be given, we simply 
note that the stochastic delay becomes ineffective at high car densities
and the system behaves increasingly as a deterministic model in the high  
density regime.  
Moreover, the proof in Refs.[29,30] works both for the deterministic model  
corresponding to $f=0$ and for the totally delayed model with $f=1$ 
corresponding to a deterministic model with $M-1$.  The 
existence of stochastic delays will only affect the time for the traffic 
to approach the asymptotic limit from its initial configuration, but 
not the nature of the asymptotic steady state.

\vspace*{0.15 true in}
\begin{center}
{\bf V. LOW DENSITY CASE}
\end{center}

For low car densities ($\rho \leq 1/M$) and $M\geq 2$, 
the average inter-car spacing
is $\bar C=1/\rho -1\geq M-1$.  It can be shown that in the
asymptotic steady state, every spacing 
will not be shorter than $M-1$, i.e., 
\begin{equation}
C_n\left( t\right) \geq M-1,\qquad \forall n
\end{equation}
or equivalently,
$N_0=N_1=...=N_{M-2}=0$, where $N_{m}$ is the number of cars 
with $m$ empty sites ahead.  
For the deterministic FI model with $v_{max}$ =$M$, 
it has been proven$^{30}$ that the steady state corresponding to 
$C_n\left( t\right) \geq M$ for all $n$ in the low car density regime with 
$\rho \leq \frac 1{V_{\max }+1}=\frac 1{M+1}$ is approached after 
a finite period of time.  
As both the $f=0$ and $f=1$ limits of the generalized FI model correspond
to deterministic FI models with $v_{max} = M$ and $v_{max} = M-1$, 
respectively, the inequality in (17) holds.  Thus, in the steady state, 
\begin{equation}
P_0=P_1=...=P_{M-2}=0. 
\end{equation}
It follows that 
$P_{long}=1-P_{M-1}$, and the average speed in the steady state (Eq.(5)) 
can be written as 
\begin{equation}
V = (M-1)P_{M-1} + (M-f) P_{long} = (M-f) - (1-f) P_{M-1}. 
\end{equation}
Thus the problem of finding $V$ amounts to obtaining $P_{M-1}$ in the 
asymptotic limit. 

To obtain $P_j$ in the steady state, we introduce 
$N_{j\rightarrow j\pm 1}$ to 
describe the number of inter-car spacings 
with their lengths changed from $j$ at time
$t$ to $j\pm 1$ at time $t+1$. 
The probability of finding an inter-car spacing with length $j$ 
at time $t$ and length $j \pm 1$ at time $t+1$ is
\begin{equation}
W_{j\rightarrow j\pm 1}\left( t\right) \equiv N_{j\rightarrow j\pm 1}\left(
t\right) /N=(N_j\left( t\right) /N)\cdot [N_{j\rightarrow j\pm 1}(t)/N_j(t)].
\end{equation}

>From Eqs.(6) and (7), we have
\begin{equation}
N_{M-1\rightarrow M}\left( t\right) /N_{M-1}\left( t\right)
=(1-f)(P_M+P_{M+1}+ \cdots), 
\end{equation}
and 
\begin{equation}
W_{M-1\rightarrow M}\left( t\right) =\left( 1-f\right) P_{M-1}\cdot
P_{long}.
\end{equation}
Similarly, from Eqs.(10) and (11), we have
\begin{equation}
N_{M\rightarrow M-1}\left( t\right) /N_M\left( t\right) =\left( 1-f\right)
P_{M-1}+\left( 1-f\right) f\left( P_M+P_{M+1}+\cdots\right),
\end{equation}
and
\begin{equation}
W_{M\rightarrow M-1}\left( t\right) =\left( 1-f\right)
P_M(P_{M-1}+fP_{long}).
\end{equation}

For $W_{j\rightarrow j\pm 1}$ with $j > M$, Eq.(12) states that for 
$C_{n}(t) = j >M$ and $C_{n+1}(t) = M-1$, 
\begin{eqnarray}
C_n\left( t+1\right) &=& j + M -1 - [(M-1)\theta_{n}(f) + 
M \theta_{n}(1-f)] \nonumber \\
&=&
j\theta _n\left( f\right) +(j-1)\theta _n\left(
1-f\right). 
\end{eqnarray}
Similarly for $C_{n}(t) = j >M$ and 
$C_{n+1}(t)\geq M$, Eq.(13) gives 
\begin{eqnarray}
C_n(t+1)&=& j + [(M-1)\theta_{n+1}(f) + M\theta_{n+1}(1-f)] 
- [(M-1)\theta_{n}(f) + M \theta_{n}(1-f)] \nonumber \\ 
&=& (j-1)\theta _n(1-f)\theta _{n+1}(f)+(j+1)\theta _n(f)\theta
_{n+1}(1-f)\nonumber\\
&&+j[\theta _n(f)\theta _{n+1}(f)+\theta _n(1-f)\theta
_{n+1}(1-f)]. 
\end{eqnarray}
Hence
\begin{equation}
N_{j\rightarrow j-1}(t)/N_j(t)=(1-f)\cdot [P_{M-1}+f(P_M+P_{M+1}+...)], 
\end{equation}
and 
\begin{equation}
N_{j\rightarrow j+1}(t)/N_j(t)=f(1-f)\cdot (P_M+P_{M+1}+...).
\end{equation}
The probabilities $W_{j \rightarrow j\pm 1}(t)$ are then given by 
\begin{equation}
W_{j\rightarrow j-1}(t)=(1-f)P_j (P_{M-1}+f \cdot P_{long}), \quad j>M
\end{equation}
and 
\begin{equation}
W_{j\rightarrow j+1}(t)=f(1-f)P_j \cdot P_{long}, \quad j>M.
\end{equation}

The asymptotic steady state should satisfy the detailed balance conditions
given by 
\begin{eqnarray}
W_{M-1\rightarrow M}(t)=W_{M\rightarrow M-1}(t); \nonumber \\ 
W_{j\rightarrow
j-1}(t)=W_{j-1\rightarrow j}(t),\quad j>M. 
\end{eqnarray}
Substituting Eqs.(22), (24), (29) and (30), we obtain
\begin{equation}
\frac{P_M}{P_{M-1}}=\frac{1-P_{M-1}}{f+(1-f)P_{M-1}}, 
\end{equation}
and 
\begin{equation}
\frac{P_j}{P_{j-1}}=\frac{f(1-P_{M-1})}{P_{M-1}+f(1-P_{M-1})}\equiv 
\alpha, \qquad  j>M, 
\end{equation}
where $\alpha$ is a constant depending only on $P_{M-1}$. 
Equations (32) and (33) imply that 
\begin{equation}
P_M=(\alpha /f)P_{M-1},\quad P_{M+1}=\alpha P_M, \quad ...,\quad P_n=\alpha
^{n-M}P_M.
\end{equation}
Together with the equality 
\begin{equation}
\bar C = \frac{1}{\rho} - 1 = \sum\limits_{j=M-1}^\infty j\cdot P_j, 
\end{equation}
we obtain 
\begin{equation}
P_{M-1}=\frac{\bar C\cdot f\cdot (1-\alpha )^2}{\alpha ^2+M\cdot \alpha
(1-\alpha )+(M-1)\cdot f(1-\alpha )^2}. 
\end{equation}
Substituting Eq.(33) for the 
constant $\alpha$, we arrive at a quadratic equation
\begin{equation}
(1-f)P_{M-1}^2+(\bar C-M+2f)P_{M-1}-f=0
\end{equation}
for $P_{M-1}$ which gives a 
non-negative root
\begin{equation}
P_{M-1}=\frac{M-\bar C-2f+\sqrt{(\bar C-M+2f)^2+4f(1-f)}}{2(1-f)}. 
\end{equation}
Substituting Eq.(38) for $P_{M-1}$ back into Eq.(19) for the 
average speed in the steady state, we finally obtain 
\begin{eqnarray}
V(t\rightarrow \infty )&=&M-f-\frac{M-\bar C-2f+
\sqrt{(\bar C-M+2f)^2+4f(1-f)}%
}2\nonumber\\
   &=&\frac{M-1+1/\rho -\sqrt{(1/\rho -1-M+2f)^2+4f(1-f)}}2. 
\end{eqnarray}
Equations (16) and (39) are the main results of the present work.  They 
give the average speed of cars in the steady state over the whole 
range of car densities for arbitrary maximum velocity and 
degrees of stochastic delays. 

\vspace*{0.15 true in}
\begin{center}
{\bf VI. DISCUSSION}
\end{center}

The fundamental diagram, i.e. the speed - car-density relation, 
of the FI model with stochastic delay is 
\begin{equation}
V(t\rightarrow \infty )=\left\{\begin{array}{ll}
	       \frac{M-1+1/\rho -\sqrt{(1/\rho -1-M+2f)^2+4f(1-f)}}%
	       2\qquad  & 0\leq \rho \leq \frac 1 M\\
	       \frac 1\rho -1 \qquad &\frac 1M\leq \rho \leq 1.
	       \end{array}\right.
\end{equation}
The general features of the speed in the steady state 
are that 
for given M and in low density regime ($\rho \leq 1/M$), different values  
of $f$
correspond to separate curves; while in the high density regime ($%
\rho \geq 1/M$), the curves corresponding to different values of 
$f$ coincide with
each other and fall into one curve.  The latter feature reflects 
that stochastic delays become ineffective for sufficiently large 
car densities and systems with different values of $f$ 
behave in the same way.  As the car density increases, the
curves for different values of $f$ meet at $\rho = 1/M$. 

In Ref.[17], numerical results have been reported for the $M=2$ FI 
model without good analytic explanations.  It is thus illustrative 
to compare the present result with numerical simulations.  For $M=2$, 
our general result (Eq.(40)) reads
\begin{equation}
V(t\rightarrow \infty )=\left\{\begin{array}{ll}
\frac{1+\rho -\sqrt{(9-8f)\rho ^2-2(3-2f)\rho +1}}{%
2\rho } \qquad & 0\leq \rho \leq 1/2\\
\frac 1\rho -1 \qquad & 1/2\leq \rho \leq 1. 
		       \end{array}\right.
\end{equation}
In order to compare with the analytic result, we carried out numerical 
simulations on one-dimensional chain with $1000$ cars and the length of 
the chain was adjusted so as to give the desired density of cars.  
Periodic boundary condition was imposed.  The motion of the cars was   
followed and the average speed of cars recorded.  
The first $20000$ time steps 
were not included into the averaging procedure so as to ensure that  
the system has approached the steady state.  
The averages were taken over the next 
$80000$ time steps.  Our numerical results are consistent with those 
reported in Ref.[17].  
Figure 1 compares the analytic results with numerical results for 
different values of $f$.  Excellent agreement is found.  
Equation (41) thus complements the numerical 
results in Ref.[17] and provides an analytic expression for the 
numerical data.  

To further establish the validity of our result, we carried out 
numerical simulations for the $M=3$ FI model with stochastic delays and 
compared results with our general expression.  Results are    
shown in Fig. 2.  Again, it is obvious 
from the figure that the agreement is excellent.   In passing, we note that 
the $M=1$ results agree well with numerical data.  For $M=1$, the FI 
model is identical to the NS model with a maximum velocity of unity and 
the $M=1$ results reported here agree with those reported in the literature
for the NS model for this particular case. 

In summary, we derived exact results for the average speed in 
the asymptotic steady state as a function of the car density and 
the degree of stochastic delay for the Fukui-Ishibashi traffic 
flow model.  The approach is based on the study of the time evolution   
of the inter-car spacings.  
The notions of long
and short inter-car spacings are introduced.  
The probability of finding a long
or short spacing on the road is then calculated. 
In the high density
regime ($\rho \geq 1/M$), all inter-car spacings will become short
spacings in the steady state; while in low density regime 
($\rho \leq 1/M$), all inter-car
spacings will be longer than or equal to length $M$-1. 
The probabilities that a spacing becomes longer and shorter by  
one unit of length in a time step are calculated.  
The asymptotic steady state is obtained by imposing the condition  
of detailed balance.  
The general expression for the average speed in the steady state 
is then obtained analytically 
as a function of the car density for arbitrary value of the maximum 
velocity and arbitrary degree of stochastic delay.  Results are 
compared with numerical data for $M=2$ and $M=3$ over the whole 
range of $0 \leq f \leq 1$.
Our analytic results are in excellent 
agreement with numerical data.  

The present approach provides an alternative way to study traffic flow 
problems analytically.  In principle, our approach can be 
extended to study other models in one and two 
dimensions [3,4,9,10,13,14,21].  While it is relatively simple to study the 
time evolution of car spacings in one-dimensional models, it is 
non-trivial to extend the present approach to two-dimensional models.  
In 2D model such as the BML model [4], 
the spacings along one direction will be coupled to the 
evolution of the spacing in another direction as a group of short 
spacings in one direction will slow down the traffic flow in the 
perpendicular direction and hence influence the car spacings in the 
other direction.  This coupling leads to complicated coupled equations 
for the probabilities of having long and short spacings in the two 
directions.  Although more complicated than the one-dimensional case, 
this coupling in turns leads to the interesting phenomena of having 
a jamming to moving phase transition at a finite car density.  Work along 
this line is in progress and results will be reported elsewhere. 

\vspace*{0.15 true in}
\begin{center}
{\bf ACKNOWLEDGMENTS}
\end{center}

BHW acknowledges the support from the Chinese National Basic Research 
Climbing Project ''Nonlinear Science'', the National Natural Science 
Foundation in China.  PMH and BHW acknowledge the support 
from the Research Grant Council (RGC) of the Hong Kong SAR Government 
through the grant CUHK 4191/97P.  The work of BH and BHW was 
also supported in part by grants from the Hong Kong Research 
Grants Council (RGC) and the Hong Kong Baptist University Faculty 
Research Grants (FRG).  The authors would also like to thank 
Dr. L.H. Tang and the members of the Center for Nonlinear Studies at HKBU 
for stimulating discussions.  

\vspace*{0.15 true in}
\begin{center}
{\bf Figure Captions}
\end{center}

Figure 1: The fundamental diagram of the Fukui-Ishibashi traffic flow model 
with the maximum car velocity $M=2$ and for different values of the 
degree of stochastic delay $f$.  
The solid curves are the theoretical
results in Eq.(41).
The points with different symbols represent
results obtained by numerical simulations.  The curves from the top down 
along the average velocity axis correspond to different values of $f$ 
between $f=0$ to $f=1$ in step of $0.1$.

Figure 2: The fundamental diagram of the Fukui-Ishibashi traffic flow model
with the maximum car velocity $M=3$ and for different values of the degree
of stochastic delay $f$.  The solid curves are the theoretical results 
in Eq.(40).  The points with different symbols represent results 
obtained by numerical simulations.  The curves from the top down along 
the average velocity axis correspond to different values of $f$ 
between $f=0$ and $f=1$ in step of $0.1$.

\vspace*{0.15 true in}
\baselineskip = 10pt
\begin{center}
{\bf References}
\end{center}
\begin{itemize}
\item[1]        S. Wolfram, {\em Theory and Application of Cellular Automata},  
	(World Scientific, Singapore,1986).
\item[2]        D. E. Wolf , M. Schreckenberg, and A. Bachem, (eds.)    
	{\em Traffic and Granular Flow}, (World Scientific, Singapore 1996). 
\item[3]        K. Nagel and M. Schreckenberg,  J. Phys. I (France) 
{\bf 2}, 2221 (1992).
\item[4] O. Biham, A. A. Middleton, and D. Levine,      
	Phys. Rev. A {\bf 46}, R6124 (1992).
\item[5]        M J. Lighthill , and G. B. Whitham,  
Proc. Roy. Soc. of London A {\bf 229}, 317 (1955).
\item[6]        B. S. Kerner and P. Konhauser,
	Phys. Rev. E, {\bf 48}, R2335 (1993). 
\item[7]        D. Helbing ,  Phys. Rev. E, {\bf 55}, 3735 (1997).
\item[8]        T. Nagatani,  J. Phys. A {\bf 26}, L1015 (1993).
\item[9] S. C. Benjamin, N. F. Johnson , and P. M. Hui,
	J. Phys. A: Math. Gen. {\bf 29}, 3119 (1996).
\item[10] T. Nagatani,  Phys. Rev. E {\bf 48}, 3290 (1993).
\item[11] T. Nagatani,  Physica A {\bf 233}, 137 (1996).
\item[12] K. H. Chung and P. M. Hui,    J. Phys. Soc. Japan 
{\bf 63}, 4338 (1994).
\item[13] M. Schreckenberg A. Schadschneider, K. Nagel and N. Ito, 
	Phys. Rev. E {\bf 51}, 2939 (1995).
\item[14] T. Nagatani,   J. Phys. Soc. Japan {\bf 62}, 2656 (1993).
\item[15] K. H. Chung , P. M. Hui, and G. Q. Gu,  
Phys Rev. E {\bf 51}, 772 (1995).
\item[16] K. Nagel, ``Experiences with iterated traffic microsimulations
in Dallas", LANL preprint adap/org/9712001 and references therein.
\item[17] M. Fukui and Y. Ishibashi, J. Phys. Soc. Japan  {\bf 65}, 
1868 (1996).
\item[18] N. H. Gartener and N. H. M. Wilson, (ed.) 
	{\em Transportation and Traffic Theory}, (Elsevier, New York, 1987).
\item[19] H. F. Chau, P. M. Hui, and Y. F. Woo, J. Phys. Soc. 
Japan {\bf 64}, 3570 (1995).
\item[20] B. H. Wang , Y. F. Woo, and P. M. Hui, 
J. Phys. A: Math. Gen. {\bf 29}, L31 (1996).
\item[21] B. H. Wang, Y. F. Woo, and P. M. Hui,  
J. Phys. Soc. Japan {\bf 65}, 2345 (1996).
\item[22] A. Schadschneider and M. Schreckenberg, 
J. Phys. A: Math. Gen. {\bf 30}, L69 (1997).
\item[23] B. H. Wang , P. M. Hui, and G. Q. Gu, 
Chin. Phys. Lett. {\bf 14}, 202 (1997).
\item[24]       B. H. Wang and P. M. Hui,  
J. Phys. Soc. Japan {\bf 66}, 1238 (1997).
\item[25]       B. H. Wang , Y. R. Kwong, and P. M. Hui,
	Physica Sinica {\bf 47}, (1998) (in press). 
\item[26]       B. H. Wang, Y. R. Kwong, and P. M. Hui, 
Phys. Rev. E {\bf 57}, 2568 (1998).
\item[27]       B. H. Wang, Y. R. Kwong, and P. M. Hui, Physica A (1998) 
(in press).
\item[28] B. H. Wang, and L. Wang, Journal of Nonlinear Dynamics in Science and Technology 
	(in Chinese), {\bf 4}, 374 (1997).
\item[29] L. Wang , B. H. Wang and P. M. Hui, 
	Acta Physica Sinica (Overseas Edition) {\bf 6}, 829 (1997).
\item[30] B. H. Wang, L. Wang, and P. M. Hui, J. Phys. Soc. Japan 
{\bf 66}, 3683 (1997).
\end{itemize}

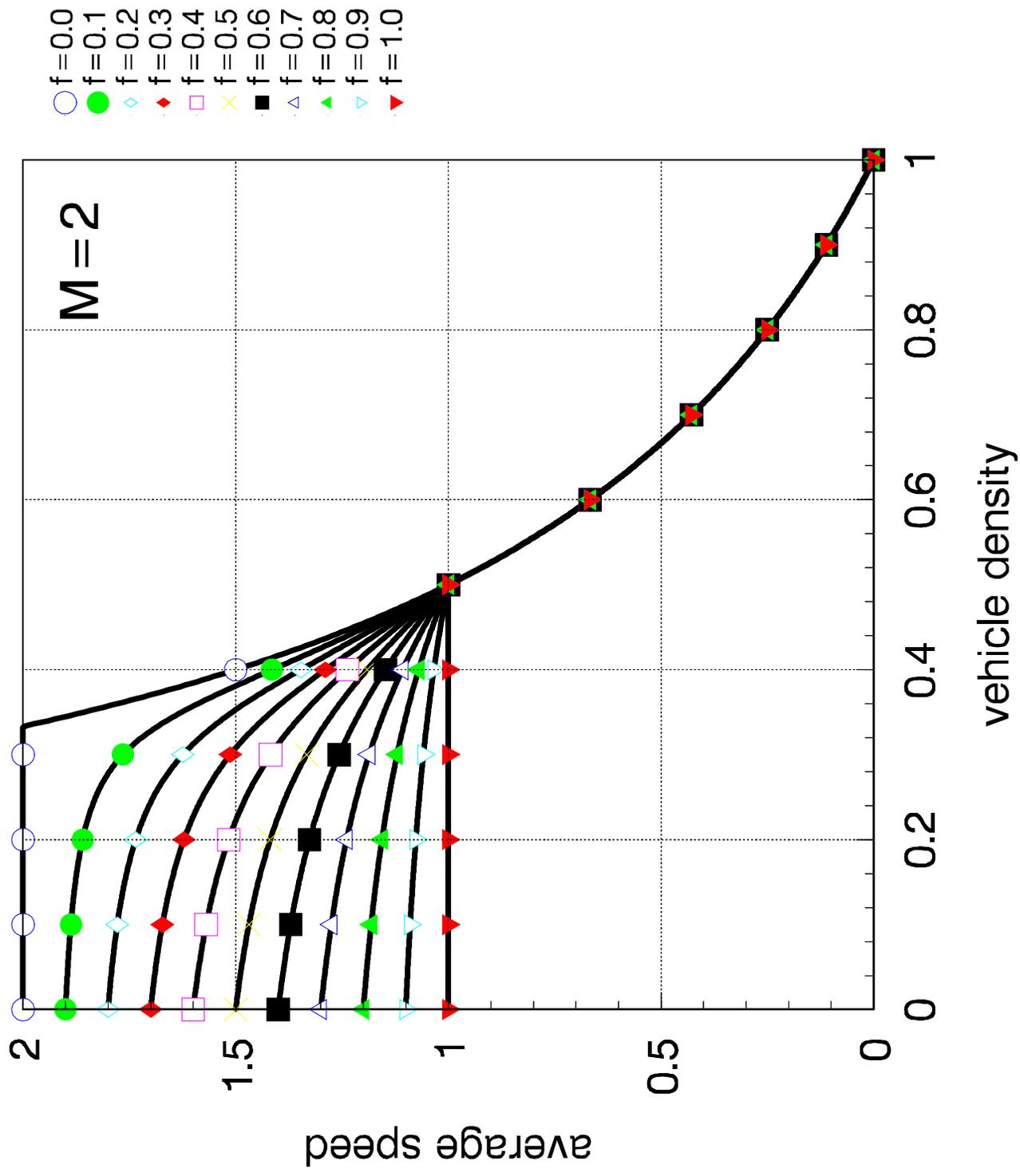
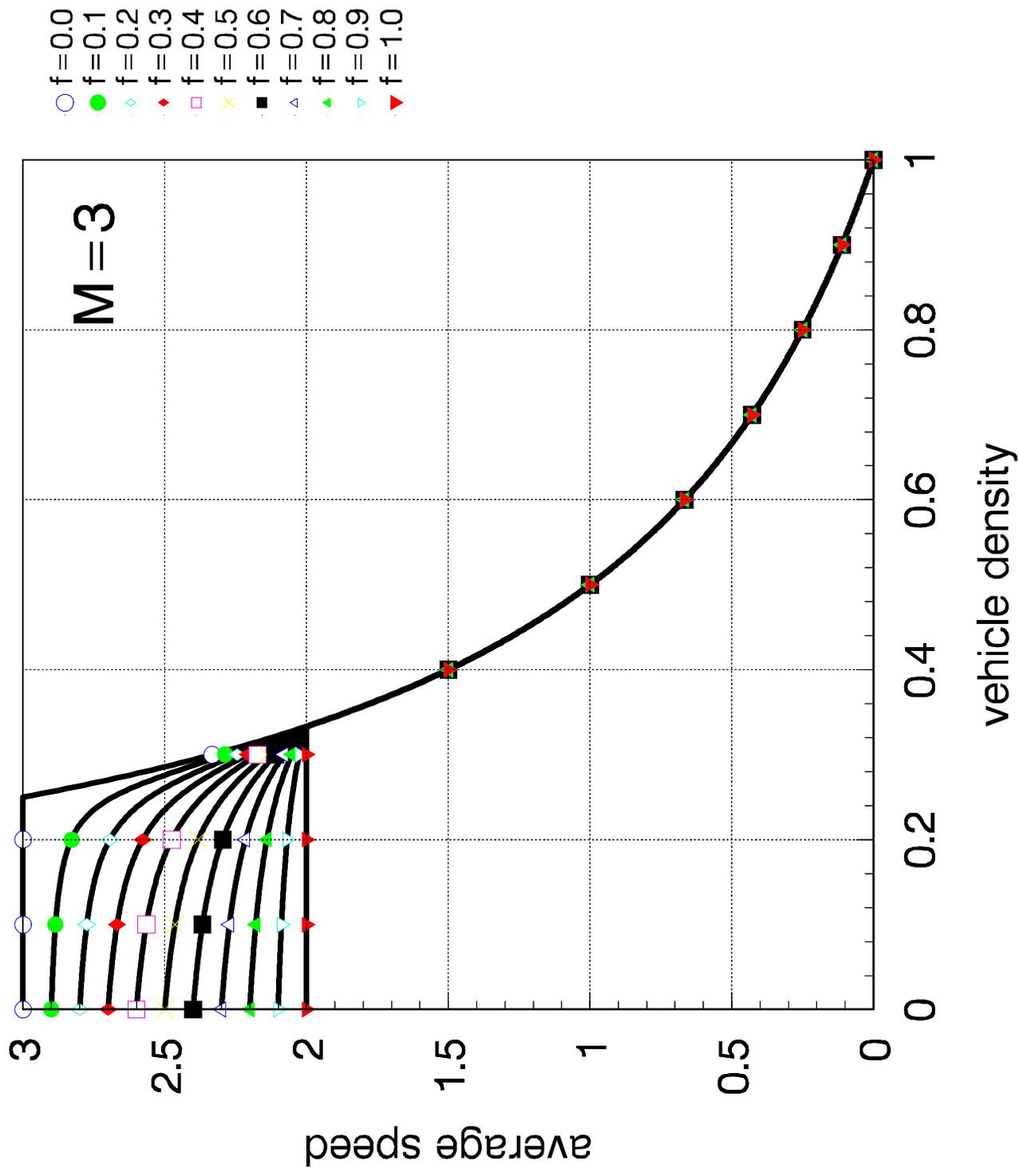
\end{document}